\documentclass[aps,prl,groupedaddress,superscriptaddress,nofootinbib, twocolumn, floatfix]{revtex4-2}

\usepackage{hyperref}
\usepackage{amsmath}
\usepackage{amsfonts}
\usepackage{amssymb}
\usepackage{graphicx}
\usepackage[version=4]{mhchem}

\usepackage{pgfplots}
\pgfplotsset{compat=1.17}

\begin{document}

\title{Non-Equilibrium Statistical Physics Beyond the Ideal Heat Bath Approximation}

\author{Jonathan Asher Pachter}
\email{jonathan.pachter@stonybrook.edu}
\affiliation{Laufer Center for Physical and Quantitative Biology, Stony Brook University, Stony Brook, NY 11794}
\affiliation{Department of Physics and Astronomy, Stony Brook University, Stony Brook, NY 11794}

\author{Ken A Dill}
\email{dill@laufercenter.org}
\affiliation{Laufer Center for Physical and Quantitative Biology, Stony Brook University, Stony Brook, NY 11794}
\affiliation{Department of Physics and Astronomy, Stony Brook University, Stony Brook, NY 11794}
\affiliation{Department of Chemistry, Stony Brook University, Stony Brook, NY 11794}

\begin{abstract}
    
Important models of nonequilibrium statistical physics (NESP) are limited by a commonly used, but often unrecognized, near-equilibrium approximation.  Fokker-Planck and Langevin equations, the Einstein and random-flight diffusion models, and the Schnakenberg model of biochemical networks suppose that fluctuations are due to an ideal equilibrium bath.  But far from equilibrium, this perfect bath concept does not hold.  A more principled approach should derive the rate fluctuations from an underlying dynamical model, rather than assuming a particular form.  Here, using \textit{Maximum Caliber} as the underlying principle, we derive corrections for NESP processes in an imperfect - but more realistic - environment, corrections which become particularly important for a system driven strongly away from equilibrium.  Beyond characterizing a heat bath by the single equilibrium property of its temperature, the bath's speed and size must also be used to characterize the bath's ability to handle fast or large fluctuations.
\end{abstract}

\maketitle

\section*{Introduction}

Non-Equilibrium Statistical Physics (NESP) aims to explain force-flow relationships in situations where fluctuations and rate distributions matter.  However, modeling in NESP often rests on near-equilibrium approximations that we describe below.  But how do we go beyond these approximations to treat flows further from equilibrium?  We desire an NESP procedure resembling \textit{Equilibrium} Statistical Physics (ESP).  \\

In ESP, you assert: (i) the Second Law of Thermodynamics, equivalent to the Maximum Entropy (MaxEnt) variational principle; and (ii) the existence of a ``perfect bath'' holding the system at constant known average quantities, like energy or particle number. This gives a principled way to \textit{derive} the full equilibrium probability distribution.  It gives the Boltzmann Distribution (BD) law with exponential dependence on energy  - a result of great generality - which can give macroscopic behaviors from a microscopic model of energy levels. \\

In contrast, traditional modeling in NESP has a more limited foundation.  There is no consensus on the proper variational principle.  Near equilibrium, the Second Law of Thermodynamics is rooted in experiments through the Clausius relationship between entropy and heat, but there is no equivalent experimental anchor further from equilibrium.  While some have favored using the entropy dissipation rate $dS/dt$, it has been shown that this is limited to near-equilibria \cite{doi:10.1146/annurev-physchem-071119-040206}. Furthermore, common NESP models \emph{assume} near-equilibrium rate distributions, based on equilibrium temperature baths, rather than deriving them based on information pertinent to the system at hand, such as properties of flows.  \\

Unlike in ESP, traditional NESP gives no principled connection between macroscopic rate laws (e.g. Fick's, Fourier's, and Ohm's Laws, Newtonian viscosities, etc.) and microscopic modelling (e.g. the Langevin and Fokker-Planck equations, random walks and random flights, the Einstein diffusion relation, the Schnakenberg relations for biochemical networks, etc.).  In forming the latter set, fluctuations are ascribed \emph{as if} they are provided by an infinite equilibrium heat bath, but transport phenomena result from gradients, so it does not make sense to derive their fluctuations from the zero-gradient limit that an equilibrium bath supplies.  In Langevin modeling, appending Gaussian white noise is known to cause serious issues when dynamics has nonlinearities, such as in viscoelastic materials, or in the van Kampen diode, wherein Langevin modeling leads to violations of conservation of energy \cite{van1981stochastic}.  \\

Although an infinite equilibrium bath is a good approximation in many circumstances, strongly driven systems will see a bath that is finite in size and responsiveness.  If a system has large fluctuations, a finite bath will not be able to absorb or accommodate them; if a system changes too quickly, it may generate corresponding heat faster than the bath can take it up, and the bath itself will be pushed out of equilibrium.  Quantitatively, we study two dimensionless metrics for how far a bath is from its idealization: the ratio of heat capacity of the system to that of the bath, and a measure of the bath's speed in responding to any perturbation, relative to the system's speed in generating it. \\

In order to model these effects, we require a generative principle of the type indicated above as desirable: a sound variational principle that derives rate distributions from microscopic models.  It has been shown that such a principle for nonequilibria is Maximum Caliber (MaxCal) \cite{RevModPhys.85.1115, doi:10.1063/1.5012990, doi:10.1146/annurev-physchem-071119-040206}.  MaxCal is essentially the Maximum Entropy inference procedure applied to path probabilities of flow trajectories; its validity and power rest on the mathematical bedrock of the Shore and Johnson arguments regarding MaxEnt \cite{1056144}, which have been recently updated and further explained by Ariel Caticha, who constructs the entire procedure of entropic inference from the ground up based on physically meaningful requirements \cite{e23070895}.  In this paper, we employ MaxEnt and MaxCal to derive first-order corrections to baths that are not infinite in size or speed.  \\

It is most illuminating to focus on a particular model.  Here, we focus on the Fokker-Planck (FP) evolution equation for dynamics on continuous microstates:

\begin{equation} \label{eq7}
    \frac{\partial p(x,t)}{\partial t} = D \beta \frac{\partial}{\partial x} [E'(x) p(x,t)] + D \frac{\partial^2 p(x,t)}{\partial x^2}
\end{equation}
where $D$ is the diffusion coefficient for the microstates labeled by $x$, $\beta$ is the inverse temperature, and $E(x)$ is the energy landscape\footnote{$x$ here does not specifically refer to position, but rather to whatever labels the microstates, which could be positions, or velocities, or angles, etc.}.  The FP model is important due to its broad use in many disciplines, including prominent use in Stochastic Thermodynamics\footnote{Although much of the Stochastic Thermodynamics literature assumes ideal bath behavior when postulating models - see e.g. the review by Seifert \cite{Seifert_2012} - the powerful tools used to analyze the irreversibility of those models in the Stochastic Thermodynamics papers remain valid and helpful for any model, including the models in the presence of non-ideal baths that we discuss here - see e.g. the work of Yang and Qian \cite{2021JSP...182...46Y}} \cite{Seifert_2012}.  \\

\section*{Review of Dynamics in Ideal Thermal Reservoirs}

Before finding corrections to the BD and FP using MaxEnt and MaxCal, we will review how to use MaxEnt and MaxCal to derive the original, ideal-bath forms of the BD and FP.  In doing so, we will note where the ideal bath assumptions arise, as a basis for going beyond them later.  \\

\subsection*{Boltzmann Distribution from Maximum Entropy}

MaxEnt infers the microscopic probability distribution $P(x)$  - given certain limited information about a system - by maximizing the Boltzmann-Gibbs-Shannon entropy:

\begin{equation} \label{eq1a}
    S[P] = - \sum_x P(x) \ln \left[ \frac{P(x)}{Q(x)} \right]
\end{equation}
relative to some prior $Q(x)$, constrained by the known information\footnote{The sum in Eq. (\ref{eq1a}) above can of course be an integral for a continuous state space.}.  For example, to derive the BD, you impose knowledge of the average system energy $\langle E(x) \rangle$ through use of a Lagrange multiplier $\beta$, and vary $S[P]$ with uniform prior, to get:

\begin{equation} \label{eq2a}
    P(x) = \frac{1}{Z} \exp[-\beta E(x)]
\end{equation}
where $Z = \sum \exp [-\beta E(x)]$ is the partition function, normalizing the distribution. \\

To see the effects of finite bath size, we switch to a more general supersystem derivation.  As shown in \cite{RevModPhys.85.1115}, consider a large system-plus-bath supersystem constrained to have a fixed total energy $E_{tot}$, since this supersystem is assumed isolated.  In the absence of more information, MaxEnt says that the probability distribution over supersystem microstates is uniform over all allowed states.  Every supersystem microstate with energy $E_{tot}$ is equally probable.  Then, we sum/integrate over the degrees of freedom of the bath, giving the system distribution as:

\begin{equation} \label{eq8a}
\begin{split}
    P(x) &\propto \Omega_B \Big( E_{tot} - E(x) \Big) \\ &= \exp \Big[ \ln \Omega_B \Big( E_{tot} - E(x) \Big) \Big]
\end{split}
\end{equation}
where $\Omega_B(E_B)$ is the number of bath microstates with energy $E_B$. \\

Here is where the infinite-bath-size approximation comes in.  If we now assume that fluctuations in the system energy are very small compared to the energy in the bath, we can Taylor expand $\ln \Omega_B$ around equilibrium, using the microcanonical thermodynamic definition of inverse temperature:

\begin{equation} \label{eq9a}
    \beta = \frac{d \ln \Omega_B(E_B)}{dE_B} \Bigg|_{E_B = E_{tot} - \langle E \rangle}
\end{equation}
Dropping terms of higher order in system energy fluctuations, we can now write Eq. (\ref{eq8a}) as:

\begin{equation} \label{eq10a}
    P(x) \propto \exp \Big[ \ln \Omega_B \Big(E_{tot} - \langle E \rangle \Big) - \beta \Big( E(x) - \langle E \rangle \Big) \Big]
\end{equation}
which, once properly normalized, is the BD.  Later, we will include the next term in the Taylor expansion of $\ln \Omega_B$ in order to more accurately describe a system in contact with a large-but-finite heat bath.

\subsection*{Fokker-Planck Equation from Maximum Caliber}

Having reviewed the equilibrium properties of a system in an ideal thermal bath, we turn to the dynamics of such a system.  In this section, we derive the FP equation using MaxCal.  We seek the temporal evolution of a probability distribution $p(x,t)$:

\begin{equation} \label{eq12a}
\begin{split}
    &\frac{\partial p(x,t)}{\partial t} = \lim_{\epsilon \rightarrow 0} \frac{1}{\epsilon} \Big[p(x, t+\epsilon) - p(x,t) \Big] \\
    &= \lim_{\epsilon \rightarrow 0} \frac{1}{\epsilon} \left[ \int dx' \ p(x', t) \ \frac{P(x', t \ ; \ x, t+\epsilon)}{P(x')} - p(x,t) \right]
\end{split}
\end{equation}
where we have written the transition probability for a short time-step $\epsilon$ in terms of $P(x', t \ ; \ x, t+\epsilon)$ - the equilibrium probability to follow a trajectory from state $x'$ at time $t$ to state $x$ at time $t+\epsilon$ - and $P(x)$, the equilibrium state probability distribution.  What remains is to determine the form of $P(x', t \ ; \ x, t+\epsilon)$ to first order in $\epsilon$, which calls for the use of MaxCal.  \\

There are three relevant constraints: (i) knowledge of the equilibrium probability $P(x) = \int dx' \ P(x, t \ ; \ x', t + \epsilon)$ for all values of $x$, which we will impose with Lagrange multiplier $\mu(x)$; (ii) microscopic reversibility $P(x', t \ ; \ x, t + \epsilon) = P(x, t \ ; \ x', t + \epsilon)$ for all $x$ and $x'$, which we will impose with Lagrange multiplier $\alpha(x,x')$; and (iii) a diffusion constraint $\int dx \ dx' \ P(x', t \ ; \ x, t + \epsilon) (x-x')^2 = 2 D \epsilon$, ensuring the continuity of motion, which we will impose with Lagrange multiplier $\gamma$.  Collecting this information, the un-maximized Caliber functional is\footnote{The time coordinates are omitted during this derivation for the purpose of more concise notation.}:

\begin{equation} \label{eq39a}
\begin{split}
    \mathcal{C} &= - \int dx \int dx' \ P(x,x') \ln \Bigg[\frac{P(x,x')}{1/\mathcal{N}}\Bigg] \\
    &+ \int dx \ \mu(x) \left[ P(x) - \int dx' \ P(x,x')\right] \\
    &+ \int dx \int dx' \ \alpha(x,x') \big[ P(x,x') - P(x',x) \big] \\
    &+ \gamma \left[ 2 D \epsilon - \int dx \int dx' \ P(x,x') \ (x'-x)^2 \right]
\end{split}
\end{equation}
where $1/\mathcal{N}$ is the uniform prior, and we note that only the anti-symmetric part of $\alpha(x,x')$ will matter, so without loss of generality we just take it to be anti-symmetric.  Now we vary $\mathcal{C}$ with respect to the probabilities and set this variation equal to zero in order to derive the optimal expression for the probabilities $P(x,x')$:

\begin{equation} \label{eq40a}
\begin{split}
    0 &= \frac{\delta \mathcal{C}}{\delta P(x,x')} = - \ln \left[\frac{P(x,x')}{1/\mathcal{N}}\right] - 1 \\
    &- \mu(x) + 2 \alpha(x,x') - \gamma (x'-x)^2 \\
    &\Rightarrow P(x,x') =  \frac{e^{-1}e^{-\mu(x)}e^{2\alpha(x,x')} e^{-\gamma (x'-x)^2}}{\mathcal{N}}
\end{split}
\end{equation}
\\

Now, we apply constraints in order to elucidate the connections between the Lagrange multipliers and physical quantities.  First, the symmetry/reversibility constraint easily eliminates the anti-symmetric function $\alpha(x,x')$:

\begin{equation} \label{eq41a}
\begin{split}
    & P(x,x') = P(x',x) \Rightarrow 2 \alpha(x,x') = \frac{1}{2}[\mu(x) - \mu(x')] \\
    &\Rightarrow P(x,x') = R(x) R(x') \sqrt{\frac{\gamma}{\pi}} e^{-\gamma (x'-x)^2}
\end{split}
\end{equation}
where we have absorbed some constants and functions into the new function $R(x)$ without loss of generality, and we now see a normalized Gaussian for $x'-x$ contained in $P(x,x')$.  For a general time span $\epsilon$, it might be difficult to proceed, but since we are taking $\epsilon$ to zero, we know from the diffusion constraint that only very very small values of $x'-x$ should be accessible with any reasonable probability, and thus taking $\epsilon$ to zero is equivalent to taking $\gamma$ to infinity.  Thus we will make use of the singular expansion of a normalized Gaussian around zero variance, namely:

\begin{equation} \label{eq42a}
    \sqrt{\frac{\gamma}{\pi}} e^{-\gamma (x'-x)^2} = \delta(x'-x) + \frac{1}{4\gamma} \delta''(x'-x) + O(1/\gamma^2)
\end{equation}
where we will not need terms beyond the first two.  This crucial observation allows us to easily incorporate the diffusion constraint for general $P(x)$; integrating $P(x,x')$ over $x'$ using the expansion of Eq. (\ref{eq42a}) yields an expression for $R(x)$ in terms of $P(x)$, to zeroth and first orders in $1/\gamma$.  Then, the diffusion constraint yields the relation $\gamma = 1/4 D \epsilon$ as $\epsilon$ goes to zero.  Eq. (\ref{eq41a}) becomes:

\begin{equation} \label{eq14a}
\begin{split}
    &P(x', t \ ; \ x, t+\epsilon) \\ = &P(x') \delta(x-x') - \epsilon D P'(x') \delta'(x-x') \\ + &\epsilon D P(x') \delta''(x-x') + O(\epsilon^2)
\end{split}
\end{equation}
Plugging this back into Eq. (\ref{eq12a}) yields:

\begin{equation} \label{eq15a}
    \frac{\partial p(x,t)}{\partial t} = - D \frac{\partial}{\partial x} \left[p(x,t) \frac{\partial}{\partial x} \ln P(x) \right] + D \frac{\partial^2 p(x,t)}{\partial x^2}
\end{equation}
which completes the MaxCal derivation of the FP. \\

It is important to note how this derivation differs from previous derivations of the FP \cite{van1981stochastic}.  Traditionally, one begins with a master equation and then \emph{assumes} the transition rates are very narrowly peaked Gaussians, possibly employing the Kramers-Moyal expansion and the Pawula Theorem - the FP then follows.  Here, however, the transition rates are \emph{derived} to be narrowly peaked Gaussians - using the MaxCal procedure with the diffusion constraint. \\

Note that substituting the BD (Eq. \ref{eq2a}) into Eq. (\ref{eq15a}) yields Eq. (\ref{eq7}), the well-known form of the FP equation in a thermal bath.  Below, we will discuss how to extend this derivation to incorporate a dynamical fluctuating local bath, albeit one which fluctuates much more quickly than the system, so that the system sees the bath as effectively almost stationary.  For the derivation below, we will need to fully understand the appearance of timescales in our setup.  \\

Although we know that it sets the timescale for a system, the diffusion coefficient contains non-temporal units - it has dimensions of $[system \ dimension]^2 / [time]$.  Wishing to extract just the timescale, we will divide the diffusion coefficient by an appropriate squared system dimension; specifically, we define here the time constant $\kappa$ associated with a diffusion coefficient $D$ as $\kappa = D/ \sigma^2$, where $\sigma^2$ is the equilibrium variance of the system variable.  Other choices exist, but this has a clear interpretation as the inverse timescale for the system to relax to equilibrium from a perturbation.  Furthermore, the precise definition of time constant is not so important as long as we use the same definition when comparing two different systems, which we will be doing below with our system of interest and the local bath surrounding it.  It will become evident that the effects of the dynamical bath become important when the time constant for the bath is non-negligible compared to the time constant for the system.

\section*{Beyond Ideal Thermal Reservoirs with Maximum Caliber}

\subsection*{Limitation (A): Infinite-Sized Heat Bath}

In the previous section, we derived Eq. (\ref{eq8a}), the equilibrium distribution for a system in contact with a generic bath, about which we have no specific information.  Performing a Taylor expansion of $\ln \Omega_B$ for fluctuations in the system energy that are small compared to the bath energy, we arrived at Eq. (\ref{eq10a}) keeping only the leading order term, which is equivalent to the BD.  What if we include the next term, the term of order $\Big[ E(x)-\langle E \rangle \Big]^2$?  Then we get a slight variation of the BD:

\begin{equation} \label{eq16a}
\begin{split}
    &P(x) \propto \exp \Bigg( \ln \Omega_B(E_{tot} - \langle E \rangle) \\ &- \beta [E(x)-\langle E \rangle] - \frac{\beta^2}{2C_B(\beta)} [E(x)- \langle E \rangle]^2 \Bigg)
\end{split}
\end{equation}
in which the variance of the energy fluctuations has also been constrained, where $C_B(\beta) = E_B'(T=1/ \beta)$ is the (microcanonical) heat capacity of the bath at inverse temperature $\beta$\footnote{The form of Eq. (\ref{eq16a}) has appeared previously in the literature, see e.g. \cite{PhysRevE.98.020103} or \cite{PhysRevE.80.031145}.}. \\

Thus, for a system in contact with a finite-sized heat bath, the equilibrium probability distribution is no longer the simple BD, with its characteristic exponential energy dependence.  For a very-large-but-finite bath, there is a quadratic correction:

\begin{equation} \label{eq16}
\begin{split}
    P(x) \propto e^{-\beta E(x)} \left[ 1-\frac{s}{2} \left( \frac{[E(x)- \langle E \rangle]^2}{\sigma^2_E} - 1 \right) + O(s^2) \right]
\end{split}
\end{equation}
where we introduce the coefficient $s=C_S/C_B$ - the ratio of system heat capacity ($C_S$) to bath heat capacity ($C_B$) - and $\sigma^2_E$ is the variance of the system energy.  Note that - although our derivation is physically illuminating - Eq. (\ref{eq16}) could be derived directly from MaxEnt with constraints on both the average energy and energy variance of the system, so even if we cannot directly measure the heat capacity of the environment, the measured energy variance of the system can determine this correction.  Eq. (\ref{eq16}) describes the effects of a finite bath on equilibria \cite{PhysRevE.98.020103, PhysRevE.80.031145}.  \\

Now, what is the effect of a finite bath on \textit{nonequilibria}?  As discussed above, for a system relaxing to the BD, we recover the familiar gradient of the energy landscape drift term in the FP.  However, in a large-but-finite heat bath, to first order in $s$, substituting Eq. (\ref{eq16}) into Eq. (\ref{eq15a}) with $D=\kappa \sigma^2$ - as explained above - yields:

\begin{equation} \label{eq18}
\begin{split}
    &\frac{\partial p(x,t)}{\partial t} 
    = \kappa \sigma^2 \frac{\partial^2 p(x,t)}{\partial x^2} \\ &+ \kappa \sigma^2 \beta \frac{\partial}{\partial x} \left[ E'(x) \left( 1 + s \frac{[E(x) - \langle E \rangle]}{\beta \sigma^2_E} \right) p(x,t) \right]
    \end{split}
\end{equation}
This can be interpreted as an FP equation with a microstate-dependent temperature perturbation. \\

Consider one-dimensional Brownian motion: we study the velocity $v$ for a particle of mass $m$ diffusing with coefficient $D = \kappa \sigma^2 = \kappa/\beta m$ through a heat bath of inverse temperature $\beta$ and size ratio $s$.  The energy is $E(v) = \frac{1}{2} m v^2$, and we add a constant force term $-(F/m)p'(v)$, i.e. we are pulling the particle with constant force $F$.  The equilibrium probability distribution with an infinite bath is a Gaussian with variance $1/\beta m$ and zero mean.  Furthermore, the non-equilibrium steady state (NESS) with an infinite bath is a Gaussian having the same variance, but a different mean value, $\langle v \rangle_F = F/m\kappa$.  Physically, an infinite bath implies a perfectly linear drag force with fluctuations that are always symmetric about the mean and independent of the driving force.  \\

\begin{figure}
    \centering
    \includegraphics[width=0.4\textwidth]{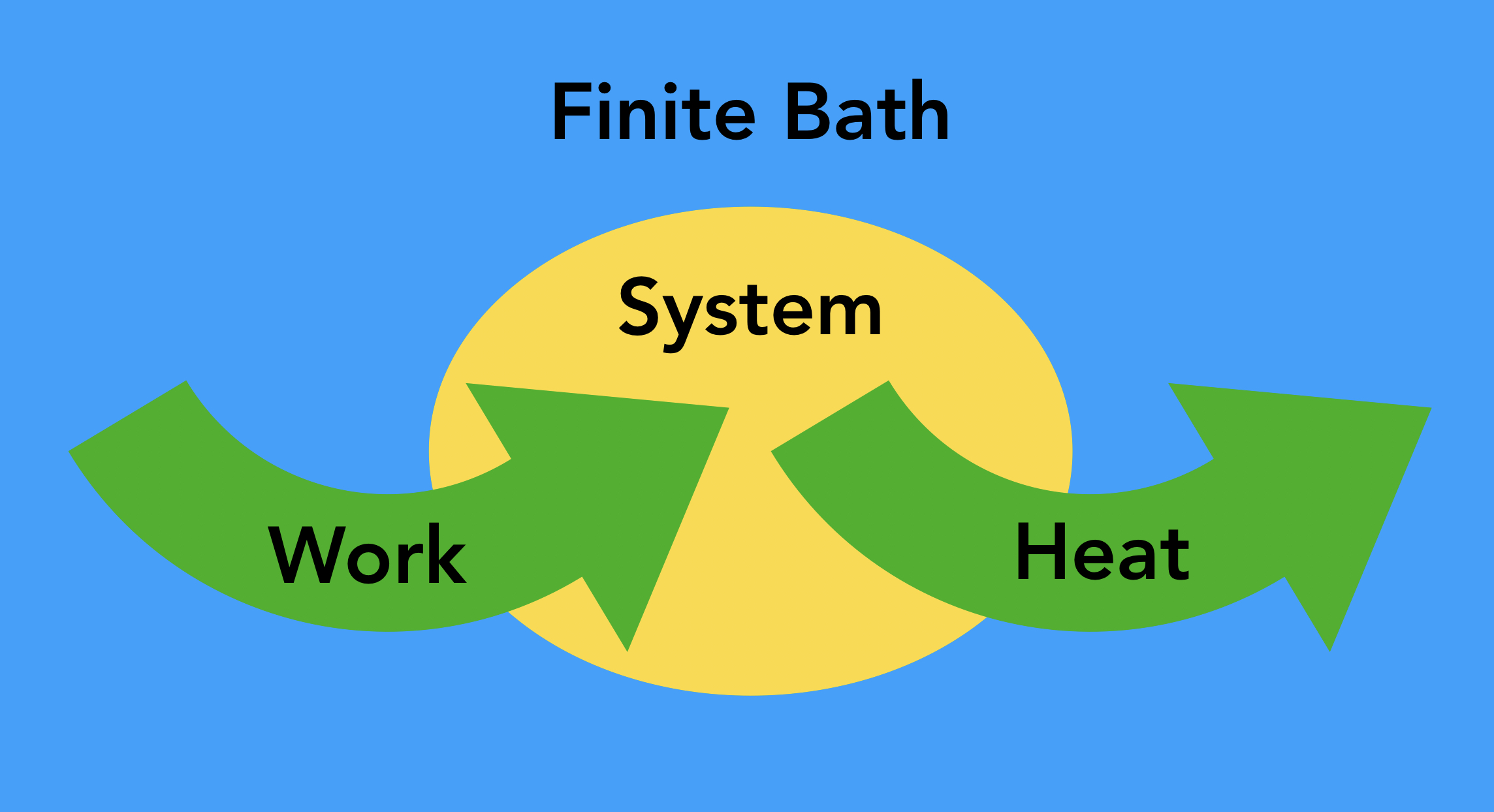}
    \caption{Energy Cycling in Finite-Bath Steady-State}
    \label{figa}
\end{figure}

As an aside, we note a troublesome aspect of the concept of a finite bath for NESS.  If the bath is finite, an externally driven process will eventually drain or saturate the bath; the system can never truly be stationary, since the bath properties will be changing constantly.  We will consider here only sufficiently large finite baths, in which there can be a relatively long transient period resembling a NESS\footnote{Alternatively, a true finite bath NESS can be achieved by an external agent cycling energy continuously between the system and the bath, rather than simply supplying it, see Fig. \ref{figa}.}. \\

\begin{figure}[ht]
\resizebox{0.23\textwidth}{!}{
\begin{tikzpicture}[every node/.append style={font=\Huge}]
\begin{axis}[legend pos = north west, xtick={0,10}, ytick={0,10}]
\addplot[dashed, red, domain=0:10, line width=4pt]{x};
\label{fig4a}
\addlegendentry{$O(s^0)$}
\addplot[green, domain=0:10, line width=4pt]{x - (x/100)*(2 + x^2)};
\addlegendentry{$O(s^1)$}
\end{axis}
\end{tikzpicture}}
\resizebox{0.23\textwidth}{!}{
\begin{tikzpicture}[every node/.append style={font=\Huge}]
\begin{axis}[legend pos = south west, xtick={0,10}, ytick={0.7,1}]
\addplot[dashed, red, domain=0:10, line width=4pt]{1};
\label{fig4b}
\addlegendentry{$O(s^0)$}
\addplot[green, domain=0:10, line width=4pt]{1 - (1/300)*(6 + x^2)};
\addlegendentry{$O(s^1)$}
\end{axis}
\end{tikzpicture}}
\caption{\raggedright{\textbf{Finite bath size: effect of large fluctuations.}  (Left:) Mean velocity vs. driving force for $s=1/100$. (Right:) Variance of velocity vs. driving force for $s=1/100$.}}
\label{fig1}
\end{figure}
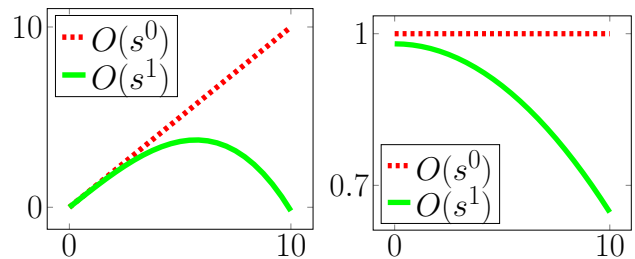

The NESS in the presence of a finite bath - i.e. the steady-state solution to Eq. (\ref{eq18}) with added external force term - is proportional to $\exp(\beta Fv/\kappa)$ times the finite-bath equilibrium distribution Eq. (\ref{eq16}).  We introduce a dimensionless version of the force $f = (F\sqrt{\beta})/(\kappa \sqrt{m})$, 
as well as absorbing a factor of $\sqrt{\beta m}$ into velocities $v$ in order to make them dimensionless.  To first order in $s$, we can write the NESS probability distribution as:

\begin{equation} \label{eq20a}
\begin{split}
    &P_f(v) = \frac{ e^{-\frac{1}{2} (v-f)^2}}{\sqrt{2 \pi}} \left(1 - \frac{s}{4} \left[ (v-f)^4 + 4f(v-f)^3 \right. \right. \\ &+ \left. \left. 2(f^2 - 2)(v-f)^2 + 4f(f^2 - 1)(v-f) - (6f^2 + 1) \right] \right)
\end{split}
\end{equation}
In this form, it is simple to calculate central moments:

\begin{equation} \label{eq24}
\begin{split}
    \langle v \rangle_f &= f - sf (2 + f^2) + O(s^2) \\
    \langle (v - \langle v \rangle_f)^2 \rangle_f &= 1 - \frac{s}{3} (6 + f^2) + O(s^2)
\end{split}
\end{equation}
and the skewness of the distribution is $-6sf + O(s^2)$.  We see nonlinear corrections to the drag force, as well as asymmetric noise which decreases in strength for stronger forcing (see Fig. \ref{fig1}). \\

One important and strange feature is that the variance in velocity seems to go negative for high enough force, which is non-physical.  This signals a phase transition, beyond which a NESS is no longer supported - when the system is driven too strongly, eventually the dissipation into the finite bath is so great that the bath cannot take it.  In an infinite-sized bath, it would never occur, but this sort of breaking-point behavior is known to appear in nature, such as when a spring is pulled too far, or in dielectric breakdown.

\subsection*{Limitation (B): Stationary Heat Bath}

\begin{figure}
    \centering
    \includegraphics[width=0.42\textwidth]{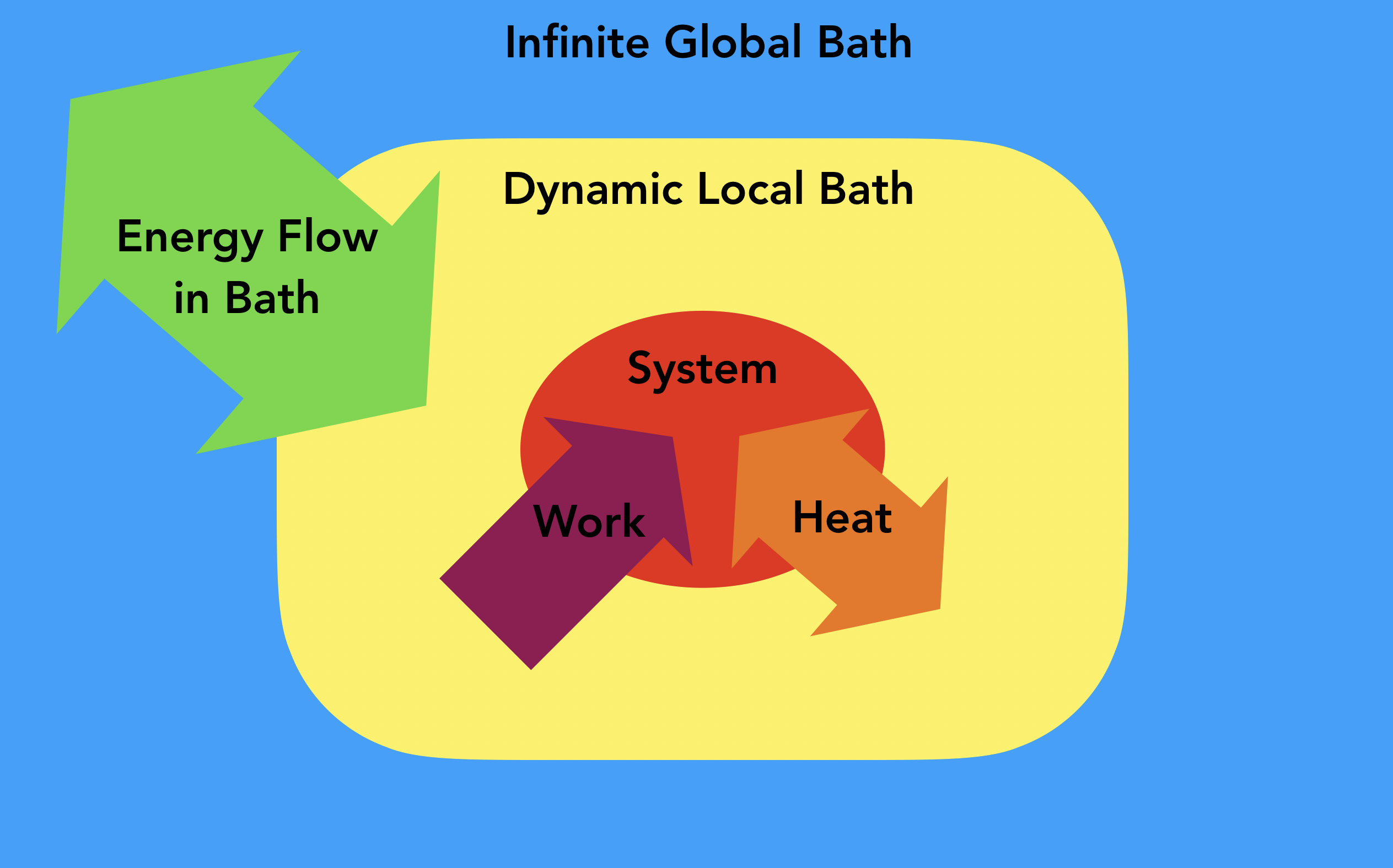}
    \caption{Energy Flow in Dynamic Local Bath Model}
    \label{figb}
\end{figure}

A second approximation for nonequilibrium models is that baths respond very rapidly to fluctuations in the system, such that they are practically stationary\footnote{By this we mean that the probability distribution of bath microstates is assumed to not change over time, even though the system may be fluctuating, because the bath is assumed to be unaffected by - i.e. uncorrelated with - the system.}.  We begin to mitigate this limitation with a framework we call the Dynamic Local Bath (DLB) model (see Fig. \ref{figb}).  The system is only coupled directly to an intermediate local bath, which itself couples to a truly infinite global bath.  This introduces a finite time constant for the environment to respond to fluctuations.  \\

In the DLB, we track not only the microstate $x$ of the system, but also one extra variable: the total energy $E_{tot}$ in the local-bath-plus-system super-system.  Since the super-system is immersed in a perfect bath, its equilibrium probability distribution is the BD:

\begin{equation} \label{eq26}
    P(x, E_{tot}) \propto \Omega_B \big( E_{tot} - E(x) \big) e^{-\beta E_{tot}}
\end{equation}
where $\Omega_B(E_B)$ is the energy degeneracy of the local bath. \\

Performing a very similar analysis to the MaxCal derivation of the FP found above, we study the dynamics of probability for the local-bath-plus-system super-system.  We use the time constant $\kappa$ associated with diffusion/fluctuations in the system, as well as introducing a new time constant $\kappa_B$ associated with diffusion/fluctuations in the super-system energy $E_{tot}$\footnote{$\kappa_B$ is directly related to the heat conductivity within the bath.}.  However, changes in $x$ as a result of interactions with the DLB are assumed uncorrelated with changes in $E_{tot}$ as a result of interactions at the far-away border between the DLB and the global bath, so there will be no cross-derivative term.  MaxCal then yields the primary evolution equation for the DLB model:

\begin{equation} \label{eq27}
\begin{split}
    &\frac{\partial p(x, E_{tot}, t)}{\partial t} \\
    &= \kappa \sigma^2 \frac{\partial}{\partial x} \big[\beta_B \big( E_{tot} - E(x) \big) E'(x) p(x, E_{tot}, t) \big] \\
    &+ \kappa \sigma^2 \frac{\partial^2 p(x, E_{tot}, t)}{\partial x^2} \\
    &+ \kappa_B \sigma_B^2 \frac{\partial}{\partial E_{tot}} \big( \big[ \beta - \beta_B \big( E_{tot} - E(x) \big) \big] p(x, E_{tot}, t) \big) \\
    &+ \kappa_B \sigma_B^2 \frac{\partial^2 p(x, E_{tot}, t)}{\partial E_{tot}^2}
\end{split}
\end{equation}
where $\sigma^2$ and $\sigma_B^2$ refer to the equilibrium variance of $x$ and $E_{tot}$, respectively, and we have defined the function $\beta_B(E_B) = \Omega_B'(E_B)$.  The resemblance to Eq. (\ref{eq15a}) - to which this equation provides a minor generalization - should be clear: there are now simply diffusion and drift terms for each variable, and the drift terms came from Eq. (\ref{eq26}), the equilibrium distribution for the super-system. \\

As a consistency check, Eq. (\ref{eq27}) reduces to Eq. (\ref{eq15a}) - the FP for the system alone - when the dimensionless response ratio $r = \kappa/\kappa_B$ goes to zero, i.e. when the local bath is perfectly stationary.  Importantly, we can go further and find corrections to Eq. (\ref{eq15a}) for an almost-but-not-perfectly-stationary bath.  To first order in $r$, we compute the expectation value:

\begin{equation} \label{eq28}
\begin{split}
    &\int dE_{tot} \ p(E_{tot}|x, t) \ \beta_B \big( E_{tot} - E(x) \big) = \beta \\ &- r \frac{\sigma^2}{\sigma_B^2} E'(x) \left[ \beta E'(x) + \frac{1}{p(x,t)} \frac{\partial p(x,t)}{\partial x} \right] + O(r^2)
\end{split}
\end{equation}
The effective temperature depends not only on the microstate $x$, but on the macrostate, since it depends on the distribution $p(x,t)$.  This correction is zero at equilibrium, i.e. when the probability distribution is the BD, and it will be small when the probability distribution is close to the BD.  \\

Using Eq. (\ref{eq28}), we integrate Eq. (\ref{eq27}) over $E_{tot}$ to find a corrected FP for fast-but-not-infinitely-responding baths, to first order in $r$:

\begin{equation} \label{eq29}
\begin{split}
    &\frac{1}{\kappa \sigma^2} \frac{\partial p(x,t)}{\partial t} = \frac{\partial^2}{\partial x^2} \left[\left( 1- r \frac{\sigma^2}{\sigma_B^2} E'(x)^2 \right) p(x,t) \right] + \\ &\frac{\partial}{\partial x} \left[ \left(\beta + r \frac{\sigma^2}{\sigma_B^2} \left[ 2E''(x) - \beta E'(x)^2 \right] \right) E'(x) p(x,t)\right]
\end{split}
\end{equation} \\

This equation is the central result of this section.  Note that previously nonexistent $x$-dependence appears in the diffusion term, slowing it down when energy flow is greater in magnitude.  Another key result is the effect on the Einstein relation, wherein the diffusion coefficient is proportional to temperature.  For Brownian motion in a DLB, we find:

\begin{equation} \label{eq30}
\begin{split}
    D = \frac{\kappa}{\beta m} \left(1 - \frac{r}{C_B(\beta)} + O(r^2) \right)
\end{split}
\end{equation}
Unless the heat capacity of the DLB $C_B(\beta)$ is independent of temperature, the diffusion coefficient will no longer be purely proportional to temperature. \\

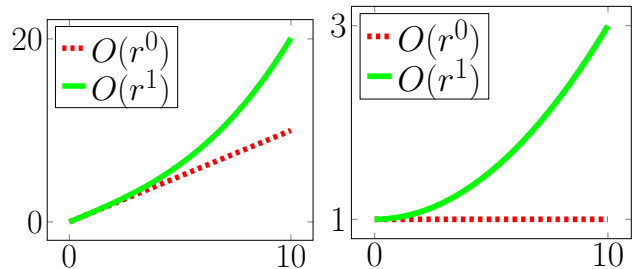
\begin{figure}[ht]
\resizebox{0.23\textwidth}{!}{
\begin{tikzpicture}[every node/.append style={font=\Huge}]
\begin{axis}[legend pos = north west, xtick={0,10}, ytick={0,20}]
\addplot[red, dashed, domain=0:10, line width=4pt]{x};
\label{fig6a}
\addlegendentry{$O(r^0)$}
\addplot[green, domain=0:10, line width=4pt]{x + (x/100)*(1 + x^2)};
\addlegendentry{$O(r^1)$}
\end{axis}
\end{tikzpicture}}
\resizebox{0.23\textwidth}{!}{
\begin{tikzpicture}[every node/.append style={font=\Huge}]
\begin{axis}[legend pos = north west, xtick={0,10}, ytick={1,3}]
\addplot[red, dashed, domain=0:10, line width=4pt]{1};
\label{fig6b}
\addlegendentry{$O(r^0)$}
\addplot[green, domain=0:10, line width=4pt]{1 + (2/100)*(x^2)};
\addlegendentry{$O(r^1)$}
\end{axis}
\end{tikzpicture}}
\caption{\raggedright{\textbf{Finite bath speed: effects of fast system fluctuations.}  (Left:) Mean velocity vs. driving force for $rs = 1/100$.  (Right:) Variance of velocity vs. driving force for $rs = 1/100$.}}
\label{fig2}
\end{figure}

Now, consider the effect of applied force.  The dimensionless velocity and force and size ratio $s=C_S/C_B$ are defined as before.  For arbitrary force $f$, we can add a force-induced drift term to Eq. (\ref{eq29}) and solve for the NESS:

\begin{equation} \label{eq31}
\begin{split}
    &P_f(v) = \frac{e^{-\frac{1}{2} (v-f)^2}}{\sqrt{2 \pi}} \left( 1 + \frac{1}{3} r s f \left[(v-f)^3 \right. \right. \\ &+ \left. \left. 3f (v-f)^2 + 3f^2 (v-f) - 3f\right] + O(r^2) \right)
\end{split}
\end{equation}
From this we can calculate the moments:

\begin{equation} \label{eq32}
\begin{split}
    \langle v \rangle_f &= f + rsf(1+f^2) + O(r^2) \\
    \langle (v- \langle v \rangle_f)^2 \rangle_f &= 1 + 2rsf^2 + O(r^2)
\end{split}    
\end{equation}
and skewness $2rsf + O(r^2)$.  Again we see nonlinear drag, as well as asymmetric noise that depends on the driving force (see Fig. \ref{fig2}).   \\

Fluctuation relations in nonequilibria, such as those of Crooks and Jarzynski, are based on ratios of forward to backward trajectory probabilities \cite{PhysRevE.60.2721,PhysRevLett.78.2690}.  In the NESS studied here, we find:

\begin{equation} \label{eq35}
    \ln \left[ \frac{P_f(v)}{P_f(-v)} \right] = 2fv + \frac{2}{3} rsf v^3 + O(r^2)
\end{equation}
This logarithmic asymmetry factor is directly proportional to dissipated power for a perfect bath, but we see a nonlinear correction for a realistic environment.  Although this is not exactly the form of the canonical Crooks formulation, we see that fluctuation relations can depend on bath speed (as well as bath size  \cite{PhysRevE.80.031145}).  \\

\section*{Summary and Conclusions: \\ The Meaning of ``Far-From-Equilibrium''}
How should we define \textit{far from equilibrium} (FFE)?  Often this means that forces and flows are large, compared to some relevant scales, and then force-flow relations are nonlinear.  But we note here two new important considerations: the ratio of system size to bath size (here quantified as $s$), to account for the bath's ability to handle large fluctuations, and the ratio of system speed to bath speed (here quantified as $r$), to account for how quickly the bath can respond to fast changes in the system.  \\

We saw that, although these quantities can have negligible effects near-equilibrium, they are increasingly important under larger driving forces.  Here, we have derived quantitative first-order corrections for nonequilibrium statistical processes in the presence of non-ideal thermal baths.  We have also extracted insight into significant qualitative differences in the example case of driven Brownian motion, including fundamentally different steady-states with breaking-point phase changes in baths of finite size, and dependence of diffusion on energy flow rate in baths of finite speed. \\

\section*{Acknowledgements}

We thank Charles Kocher, Rostam Razban, and Ying-Jen Yang for many helpful comments, and JP thanks Natalie Weiss Pachter for enlightening conversations.

\bibliography{main.bib}

\end{document}